\documentclass[10pt, a4paper]{article}
\usepackage[utf8]{inputenc}
\usepackage[T1]{fontenc}
\usepackage{graphicx}
\usepackage[colorlinks,allcolors=cyan]{hyperref}
\usepackage{amsmath,amsfonts,amssymb}
\usepackage{threeparttable}
\usepackage{authblk}
\usepackage{lipsum}

\begin{document}

\title{The Range of Cooperativity Modulates Actin Binding Protein Cluster Size, Density and Dynamics}

\author[1]{Thomas Le Goff}
\author[1]{Alph\'ee Michelot}

\affil[1]{Aix Marseille Univ, CNRS, IBDM, Turing Centre for Living Systems, Marseille, France}

\date{}

\maketitle


\begin{abstract}
The actin cytoskeleton is composed of multiple networks which are specialized for several processes such as cell motility or cell division. Each of these networks are composed of organized actin microfilaments which are decorated with specific sets of actin binding proteins (ABPs). The molecular mechanisms guiding ABPs to specific actin networks are still poorly understood, but cooperativity, the mechanism by which the binding of an ABP is positively influenced by proximal bound ABPs, plays a crucial role in generating locally dense stretches of ABPs. Cooperative binding is characterized by its amplitude, but also by the range at which its effects are propagated along an actin filament through long-range allosteric interactions. The range of these allosteric effects is still debated, but is likely to be significant at the lengthscale of actin filaments in cells. Here, we investigated how cooperativity influences the clustering of ABPs, using a stochastic computational model of binding of ABPs to actin filaments. The model reproduces the formation of ABP clusters observed experimentally at the single filament scale, and provides a theoretical estimation of the range of cooperativity for proteins such as ADF/cofilin. We found that both the amplitude and the spatial range of cooperativity dramatically impact the properties of clustering. However, the parameters of cooperativity modulate differently the rate of assembly, size and dynamics of the ABP clusters, suggesting that cooperativity is an efficient mechanism to regulate precisely the recruitment of ABPs in cells. This work provides a more general framework for future understanding of how actin networks acquire distinct and specific protein compositions from a common cytoplasm.
\end{abstract}


\section*{Introduction}
The actin cytoskeleton is an ensemble of biopolymers that cells use to power diverse functions such as cell motility or cell division. Actin filaments are not randomly organized, but assemble in a variety of structures whose properties are optimized for a given cellular function \cite{skau2015}. In particular, the organization and dynamics of the filaments is tightly controlled in time and space by specific sets of Actin Binding Proteins (ABPs). These ABPs bind to the side or to the ends of the filaments, and regulate important biochemical reactions such as their elongation rates, their crosslinking or their rate of disassembly \cite{blanchoin2014,pollard2016}. 

Importantly, individual actin networks are specifically regulated because they interact with defined sets of ABPs. While they assemble, actin networks must precisely control the accessibility of the individual filaments to the appropriate families of ABPs \cite{kovar2010,michelot2011}. As all the actin filaments of the cell are built from identical actin subunits, cells must employ efficient molecular mechanisms to specify in time and space the identity of actin filaments. As a matter of fact, if ABPs were systematically following the law of mass action, they would decorate with a similar ratio all the actin filaments of the cell. In this study, we focused on cooperativity, which for actin filaments is the property that the binding of a first ABP influences positively the attachment of other ABPs by increasing locally the association constant and/or decreasing their dissociation constant. Cooperativity represents an efficient mechanism to favor a high local density of identical ABPs, even if this family of proteins may represent only a small fraction of all the ABPs present in the cytoplasm.

The cooperative binding of proteins to the side of actin filaments can occur through two non-exclusive mechanisms. 
The first mechanism of cooperativity derives from possible interactions between two ABPs bound to consecutive sites of an actin filament. An affinity between neighboring ABPs increases the likelihood that an ABP will bind next to a previously bound one, inducing a cooperative effect. This mechanism, also called end-to-end cooperativity, is the main mechanism of cooperativity for proteins such as tropomyosins \cite{tobacman2008,christensen2017}. Tropomyosins are multi-functional coiled coil dimers that wrap around actin filaments. They bind adjacently, without gaps, because the N-terminus of one tropomyosin subunit can directly bind to the C-terminus of another tropomyosin subunit along the actin filament \cite{gunning2008}. As tropomyosin extend over several subunits of actin, end-to-end cooperativity represents an efficient mechanism to limit the existence of empty gaps on the actin filament between consecutive tropomyosins.

The second mechanism of cooperativity is independent of any potential side interaction between ABPs. Rather, a body of literature documents how actin binding proteins may also affect structurally their substrate, the actin filament \cite{galkin2001,rouiller2008,michelot2011,jensen2012}. Importantly, all the subunits of an actin filament do not adopt structural conformations independently from one another, but rather adopt preferential conformations over several actin subunits through long-range allosteric interactions \cite{orlova1995,kozuka2006,galkin2010,hild2010}. Thus, ABPs can modify locally the structure of actin filaments, which in turn favors the binding of additional ABPs and trigger cooperativity. This mechanism of cooperativity implies that cooperative binding does not only occur strictly on adjacent sites along actin filaments, but rather stochastically over the distance where the structure of the actin filament has been modified. Such mechanism of cooperativity applies for proteins such as ADF/cofilin, which are small globular proteins implicated in actin disassembly. The binding of ADF/cofilin to actin is equimolar, and induces major conformational changes on actin filaments \cite{galkin2001,cao2006,tanaka2018}. Principally, ADF/cofilin changes the mean twist of actin filaments and increases their flexibility \cite{mccullough2008,galkin2011,delacruz2009}. The cooperativity of ADF/cofilin binding to actin filaments has been reported extensively, although attempts to evaluate a range of cooperativity for ADF/cofilin along actin filaments brought considerably different results among studies. Differential scanning calorimetry experiments and spectroscopic lifetime measurements provided originally an order of magnitude of about 100 subunits for the structural changes induced by ADF/cofilin along an actin filament \cite{bobkov2006,prochniewicz2005}. Direct observations include (1) real-time fluorescence microscopy, which evaluates a range of cooperative binding of about 24 actin subunits \cite{hayakawa2014}; (2) atomic force microscopy, which revealed asymmetric conformational changes in filaments induced by ADF/cofilin propagated over one-half of a helix (i.e. about 15 subunits) \cite{ngo2015}; and (3) cryo-EM data which did not find any visible modification of the twist of actin filaments further than the first few subunits from the boundary between bare and decorated segments of the filament \cite{huehn2018}.

The uncertainty in the range of allosteric modifications of the actin filaments is problematic for our interpretation of these effects, as actin filaments are on average very short in cells. For example, in the branched network of the lamellipodium, the spacing between Y-junctions occurs within 20-50 nm, which corresponds to 8-18 subunits of actin \cite{svitkina1999}. At endocytic sites, the spacing between Y-junctions occurs within 50-200 nm, which corresponds to 18-68 subunits of actin \cite{young2004,sirotkin2010}. For linear networks, actin filaments are in general longer, for example in filopodia \cite{svitkina2003}, or in the cytokinetic ring with an average length of 0.8 $\mu$m, which corresponds to 270 actin subunits \cite{kamasaki2005}. Nevertheless, one understands easily that a range of cooperativity of 1 subunit or 100 subunits for actin filaments in cells will have major consequences for their decoration by ABPs. Moreover, very little is known about the precise repartition of ABPs at these scales in cells. 

The effect of cooperativity at the single actin filament scale was progressively unveiled with the development of real-time \textit{in vitro} fluorescence microscopy. The effect of cooperativity is obvious when the binding of ABPs is observed for various concentrations of ABPs. At a low concentration of ABP, bound molecules are isolated. At an intermediate ABP concentration, cooperative effects are most important, and ABPs begin to assemble on the side of actin filaments as clusters. The size and dynamics of these clusters appears to vary widely depending on the nature of the ABPs. At a high concentration of ABP, ABPs decorate actin filaments fully \cite{gressin2015,christensen2017}. The rapidity of these transitions depends on the amplitude of cooperativity, which can be quantified by a Hill coefficient \cite{hill1913}. However, it is not determined to this day how the amplitude and the range of cooperativity of an ABP affect the repartition of ABPs along actin filaments at the molecular level. 

For this study, our objective was to overcome current experimental limitations, by developing a stochastic model of ABP cluster formation. This model allowed us to predict the behavior of ABP cluster formation along actin filaments as function of the amplitude and the range of cooperativity. We found that our model recapitulates the experimental behavior of proteins such as ADF/cofilin, and allowed us to explore the consequences of cooperativity over a large range of parameters. Our results indicate that the amplitude and the range of cooperativity modulate extensively actin binding protein cluster size, density and dynamics. These findings suggest cooperativity as a key mechanism for cells to control the repartition of ABPs to specific actin filament populations.

\section*{Materials and Methods}

\subsection*{ADF/cofilin cluster formation and imaging \textit{in vitro}}
For the imaging of Alexa488-ADF/cofilin binding to Alexa568-labeled actin filaments, proteins were purified, labeled and experiments performed as detailed in \cite{gressin2015}. Experiments were performed in the presence of 1 $\mu$M of globular actin and 3 $\mu$M of profilin and at concentrations of ADF/cofilin for which severing events were the least frequent. Data were acquired on a Nikon Eclipse Ti microscope, equipped with a 60X NA 1,49 objective, an OptoSplit II beam splitter and a Prime 95B scientific CMOS camera (Photometrics). Images were recorded using Metamorph software, and analyzed quantitatively with ImageJ 1.49v. Average distances between clusters were evaluated by measuring the numbers of visible clusters per total length of filamentous actin in a field of view 20 minutes after the initiation of the experiment. This analysis was performed with the plugin Ridge Detection available for ImageJ. Values indicated in the text are averages from 3 independent experiments.

\subsection*{Description of the model}

To study clusters formation for a protein binding cooperatively to filamentous actin, we developed a model that we simulate with a dynamic Monte-Carlo method. The studied system corresponds to a single actin filament divided into equal compartments corresponding to sites where the ABPs can bind. If not specified, our simulations are done with an actin filament of $8.10^4$ subunits in a volume of $2.25.10^7l_{site}^3$, where $l_{site}$ is the typical length of one site. We fixed $l_{site}=10~nm$ which corresponds to the approximate size of an actin subunit. These values correspond to a concentration of $[actin]\simeq5.9~\mu M$. It models the case where the binding stoichiometry of the ABP to actin is 1:1, although results from our work could easily be interpreted by extension to other binding stoichiometries (e.g. tropomyosin which binds over the length of several actin subunits), or to any other 1 dimensional biological polymer systems interacting with side binding ligands.

The algorithm proceeds as following. Each ABP can be in two different states: attached to the actin filament or free in solution. When an ABP is bound to the filament, an energy is assigned stochastically following a Boltzmann distribution, and the ABP unbinds when its energy exceeds the activation energy of the corresponding chemical reaction. When an ABP is free in solution, the algorithm tests if the ABP is in contact with the filament. The probability of contact with the filament is equal to the ratio between the volume in which the ABP is in contact with actin and the total volume accessible to the ABP. When an ABP is in contact with the actin filament, a compartment is chosen randomly. If the compartment is already filled with another ABP, the ABP is left free in solution. On the contrary, if the compartment is empty, an energy is assigned to the ABP similarly than before and compared to the activation energy of binding. If this energy is larger than the activation energy, then the ABP binds.

We aspired to develop a model which would describe faithfully the biochemical properties of ABPs. The main parameters accessible experimentally are reaction rates $k$, which are related to activation energies $E_a$ by the Arrhenius law 
\begin{equation}
k={\cal A}\exp(-E_a/k_BT)
\end{equation}
where ${\cal A}$ is an unknown prefactor, $k_B$ is the Boltzmann constant and $T$ is the temperature. We ran a set of preliminary simulations with different values of $E_a$. The rate of ABP binding to the actin filament obtained from these simulations enabled us to compute the prefactor ${\cal A}$, and to verify that system follows the Arrhenius law (data not shown). These preliminary experiments enabled us to fix the prefactor ${\cal A}$ for the subsequent simulations.

\begin{figure}[hbt!]
\centering
\includegraphics[width=0.8\linewidth]{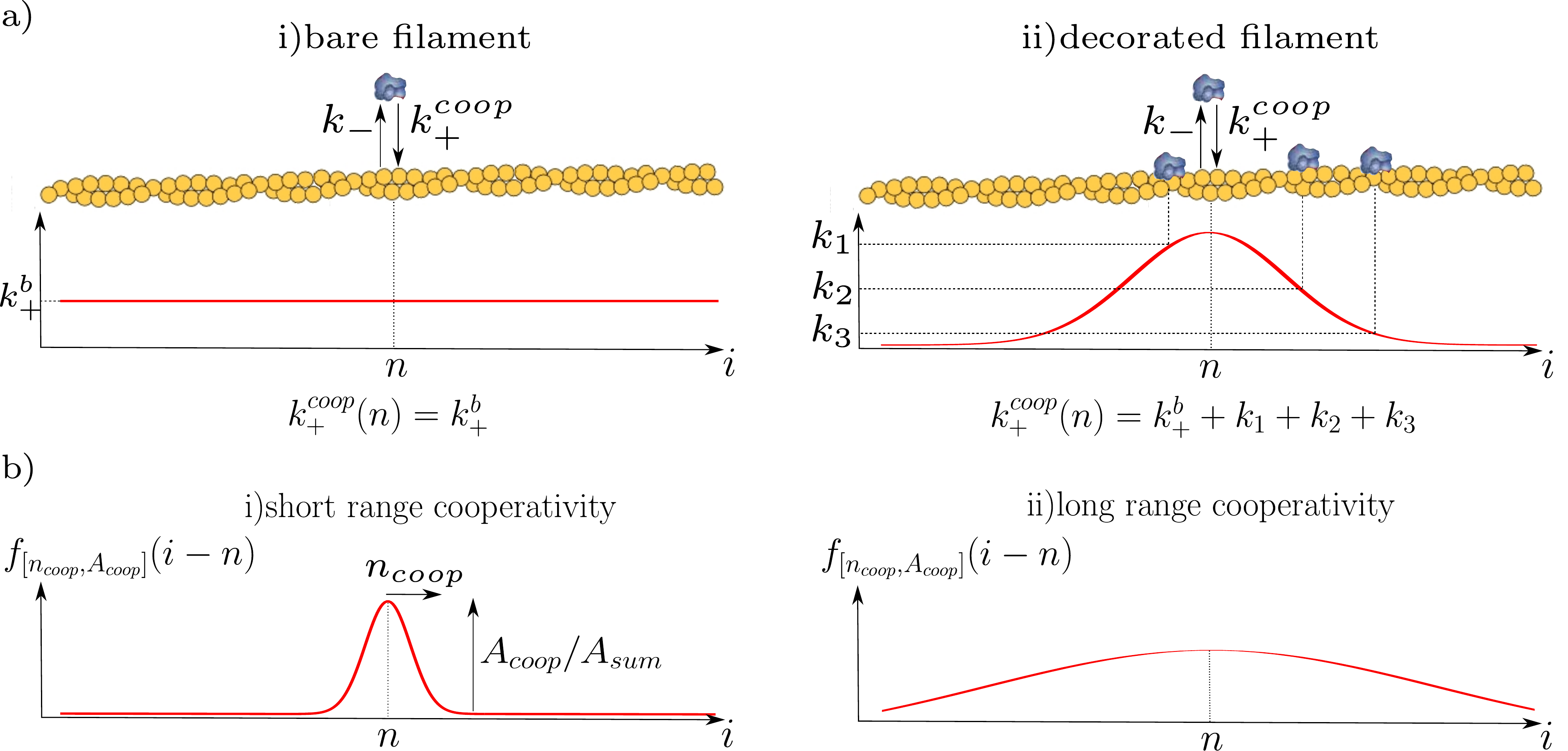}
\caption{Illustrative figure explaining the model. a) Value of binding rate on the filament : i) on a bare filament, the binding rate $k_+^{coop}$ of an ABP is constant and equal to $k_+^b$, ii) on a decorated filament, the binding rate $k_+^{coop}$ of an ABP is influenced by the presence of other bound ABPs. In both cases the dissociation rate is constant and equal to $k_-$. b) The function $f_{[n_{coop},A_{coop}]}$ defines the amplitude and range of cooperativity at the site i when a single ABP is bound at the site n. Plots give examples of i) short range and ii) long range cooperativities (see Eq.\eqref{eqfncoopacoop}).} 
\label{fig:1}
\end{figure}

In our simulations, we fixed dissociation rates $k_-$ to a constant value along the actin filament, as dissociation rates of most ABPs is not reported to be variable when ABPs dissociate spontaneously from actin filaments and was even measured to be constant for cofilin \cite{hayakawa2014}. On the contrary, binding rates of ABPs $k_+^{coop}$ take cooperativity into account in order to characterize how cooperativity may influence the dynamics and steady state regime of ABP cluster formation (Fig. \ref{fig:1}-a)). Our model enables to study the case of end-to-end cooperativity, but also allows to explore the possibility that allosteric interactions along the actin filament change the spatial range of cooperativity beyond one actin subunit along the filament \cite{bobkov2006}. In other words, binding rates $k_+^{coop}$ in our model depends on the local distribution of previously bound ABPs. The influence of cooperativity is defined by two parameters: $n_{coop}$, which quantifies the typical number of sites over which the presence of a bound ABP has an influence on binding rates; and $A_{coop}$, which is the amplitude of cooperativity (Fig. \ref{fig:1}-b)). How cooperativity is reinforced with the simultaneous binding of multiple ABPs has not been investigated experimentally to our knowledge. Therefore, we chose a model where the contribution of all surrounding bound ABPs are (see Fig.\ref{fig:1}-a)) and not only the nearest neighbor as previously studied
\cite{vilfan2001,delacruz2005,delacruz2010,christensen2017}. 

As a consequence, we consider the following binding rate
\begin{equation}
k_+^{coop}(n)=k_+^b+\sum_is(i)f_{[n_{coop},A_{coop}]}(i-n)
\label{eqkcoop}
\end{equation}
where
\begin{eqnarray}
f_{[n_{coop},A_{coop}]}(i-n)&=&A_{coop}\frac{\sum_{j\neq n}\exp\left[-\{(j-n)^2-1\}\right]}{\sum_{j\neq n}\exp\left[-\{(j-n)^2-1\}/n_{coop}^2\right]}\exp\left[-\{(i-n)^2-1\}/n_{coop}^2\right]\nonumber\\
&=&A_{coop}\frac{\exp\left[-\{(i-n)^2-1\}/n_{coop}^2\right]}{A_{sum}(n_{coop})}.
\label{eqfncoopacoop}
\end{eqnarray}
$s(i)=1$ if the compartment is occupied and $s(i)=0$ if not, $n$ is the site number, $A_{coop}$ is the amplitude of cooperativity and $k_+^b$ is the binding rate of a bare actin filament (Fig. \ref{fig:1}). $k_+^{coop}$ is defined to be, for a given $A_{coop}$, the same binding rate at the end of an infinite cluster of ABPs independently of $n_{coop}$. Most sections of the manuscript refer to the ratio $A_{coop}/k_+^b$ which normalizes the amplitude of cooperativity $A_{coop}$ by the association constant $k_+^{b}$.

\section*{Results and Discussion}

\subsection*{Validation with Hill Formalism}

We first aimed at validating our approach by comparing the results from our model with a theoretical description based on the Hill formalism. In this formalism the density of decorated actin filaments is plotted with respect to the concentration of free ABPs. Binding curves display characteristic sigmoidal shapes which represent a transition from barely decorated to fully decorated actin filaments. In the frame of the Hill model \cite{hill1913}, the steady state density of decorated actin filaments $d$ typically follows the function
\begin{equation}
d=\frac{[\text{free ABP}]^n}{K_d^{eff}+[\text{free ABP}]^n}
\label{eq:hillcoeff}
\end{equation}
where $[\text{free ABP}]$ is the concentration of free ABPs, $n$ is the Hill coefficient and $K_d^{eff}$ is an effective dissociation constant. The Eq. \eqref{eq:hillcoeff} can also be reformulated as
\begin{equation}
\frac{d}{1-d}=\frac{[\text{free ABP}]^n}{K_d^{eff}}
\label{eq:hillcoeffpowerlaw}
\end{equation}
which gives a power law and is more convenient to characterize. The Hill coefficient $n$ quantifies the degree of cooperativity and how abrupt is the transition while $K_d^{eff}$ locates this transition.

\begin{figure}[hbt!]
\centering
\includegraphics[width=1\linewidth]{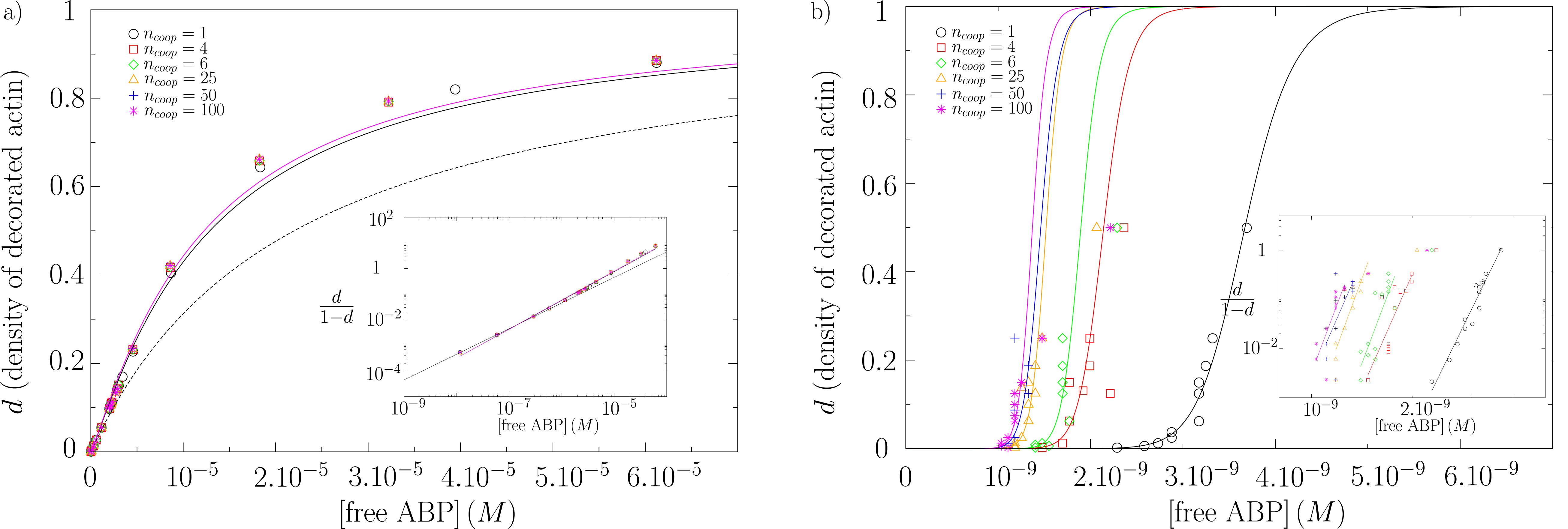}
\caption{Evolution of the steady state density of decorated actin filaments $d$ as a function of the concentration of free ABP for different values of $n_{coop}$. The symbols represent simulations results. Insets : evolution of $d/(1-d)$ in log-log scale. a) in the case of weak cooperativity ($A_{coop}/k_+^b=1$), the black solid line is the fit of Hill model for $n_{coop}=1$, the pink solid line is the fit of Hill model for $n_{coop}=100$ and the dashed line is the corresponding curve if there is no cooperativity. b) In the case of strong cooperativity ($A_{coop}/k_+^b=10^4$), the solid lines represent the different fits of Hill model. Parameters corresponding to fits of Hill model are shown in Table \ref{tab:1}.}
\label{fig:2}
\end{figure}

We fixed for these simulations the value of $K_d = 2.2.10^{-5}~M$, which is in the range of experimental values for various ABPs binding cooperatively to actin (e.g, for cofilin, $K_d\simeq10^{-6}-10^{-5}~M$ \cite{blanchoin1999,delacruz2005,delacruz2010,hayakawa2014} and for tropomyosin, $K_d\simeq10^{-6}-10^{-3}~M$ \cite{wegner1979,gateva2017,christensen2017}). We performed simulations for two values of $A_{coop}/k_+^b$, 1 and $10^4$, corresponding respectively to weak and strong cooperativities (Fig. \ref{fig:2}). For each value of $A_{coop}$, six different values of $n_{coop}$ (1, 4, 6, 25, 50 and 100) were computed. Each data point corresponds to the steady state values of $d$ and $d/(1-d)$ as a function of the concentration of free ABP in solution. 

Results obtained from the simulations confirm that the model follows closely the theory described by the Hill equation, although it diverges progressively at high concentration of ABPs for strong and long range of cooperativities (Fig. \ref{fig:2}). We can determine the Hill coefficients and the effective dissociation constants by fitting the results from our simulations with Eqs \eqref{eq:hillcoeff} and \eqref{eq:hillcoeffpowerlaw} (see Table \ref{tab:1}). In the weak cooperativity regime ($A_{coop}/k_+^b=1$), the Hill coefficient $n$ is increased of $12-14\%$ with respect to a non-cooperative situation where $n=1$. The effective dissociation constant is about $6.8-7.9$ times lower that the $K_d$ on bare actin filaments, which means an enhancement of adhesion to the filament. We note that in the case of weak cooperativity, it is rather difficult to extract a dependence with the range of cooperativity $n_{coop}$. In the strong cooperativity regime ($A_{coop}/k_+^b=10^4$), the effect is more important. The Hill coefficient is multiplied by more than ten times. The value of $n$ is now varying with the range of cooperativity $n_{coop}$. It starts from about 14 at very short range ($n_{coop}=1$) to reach about 18 for long range cooperativity. The effect on the effective dissociation constant $K_d^{eff}$ is more apparent. $K_d^{eff}$  is decreased from 113 order of magnitude at short range to about 160 order of magnitude for long range cooperativity. The evolution seems monotonous with the range of cooperativity $n_{coop}$, although the results for $n_{coop}=50$ deviates from the trend in this set of simulations. 

\begin{table}[hbt!]
\caption{Hill model's parameters obtained by fitting simulations results shown in Fig.\ref{fig:2}.}
\label{tab:1}
\centering
\begin{threeparttable}
\begin{tabular}{| c c l l | c c l l |}
\hline
$A_{coop}/k_+^b=1$ & $n_{coop}$ & $n$ & $K_d^{eff}~(M)$ & $A_{coop}/k_+^b=10^4$ & $n_{coop}$ & $n$ & $K_d^{eff}~(M)$\\\hline
  & 1   & 1.1229 & $3.227.10^{-6}$ &   & 1    & 13.922 & $3.673.10^{-118}$ \\
  & 4   & 1.1309 & $2.829.10^{-6}$ &   & 4    & 15.459 & $9.6237.10^{-135}$ \\
  & 6   & 1.138  & $2.587.10^{-6}$ &   & 6    & 18.099 & $1.366.10^{-158}$ \\
  & 25  & 1.1379 & $2.587.10^{-6}$ &   & 25   & 18.611 & $7.668.10^{-165}$ \\
  & 50  & 1.1285 & $2.866.10^{-6}$ &   & 50   & 15.667 & $3.7987.10^{-139}$ \\
  & 100 & 1.131 & $2.784.10^{-6}$ &    & 100  & 18.124 & $2.3431.10^{-161}$ \\
\hline
\end{tabular}
\end{threeparttable}
\end{table}

Overall, the behavior of the model is consistent with the Hill model, and the calculated values of Hill coefficients are consistent with the orders of magnitude that are published in the literature. The typical measured values of $n$ are in the range of $1-10$ for ADF/cofilin \cite{mccough1997,delacruz2005} and tropomyosin \cite{heald1988}. Another study \cite{moraczewska1999} suggests that $n$ could be even much larger than 10 in case of tropomyosin.

\subsection*{Impact of Cooperativity on the Length of ABP Clusters}

\begin{figure}[hbt!]
\centering
\includegraphics[width=1\linewidth]{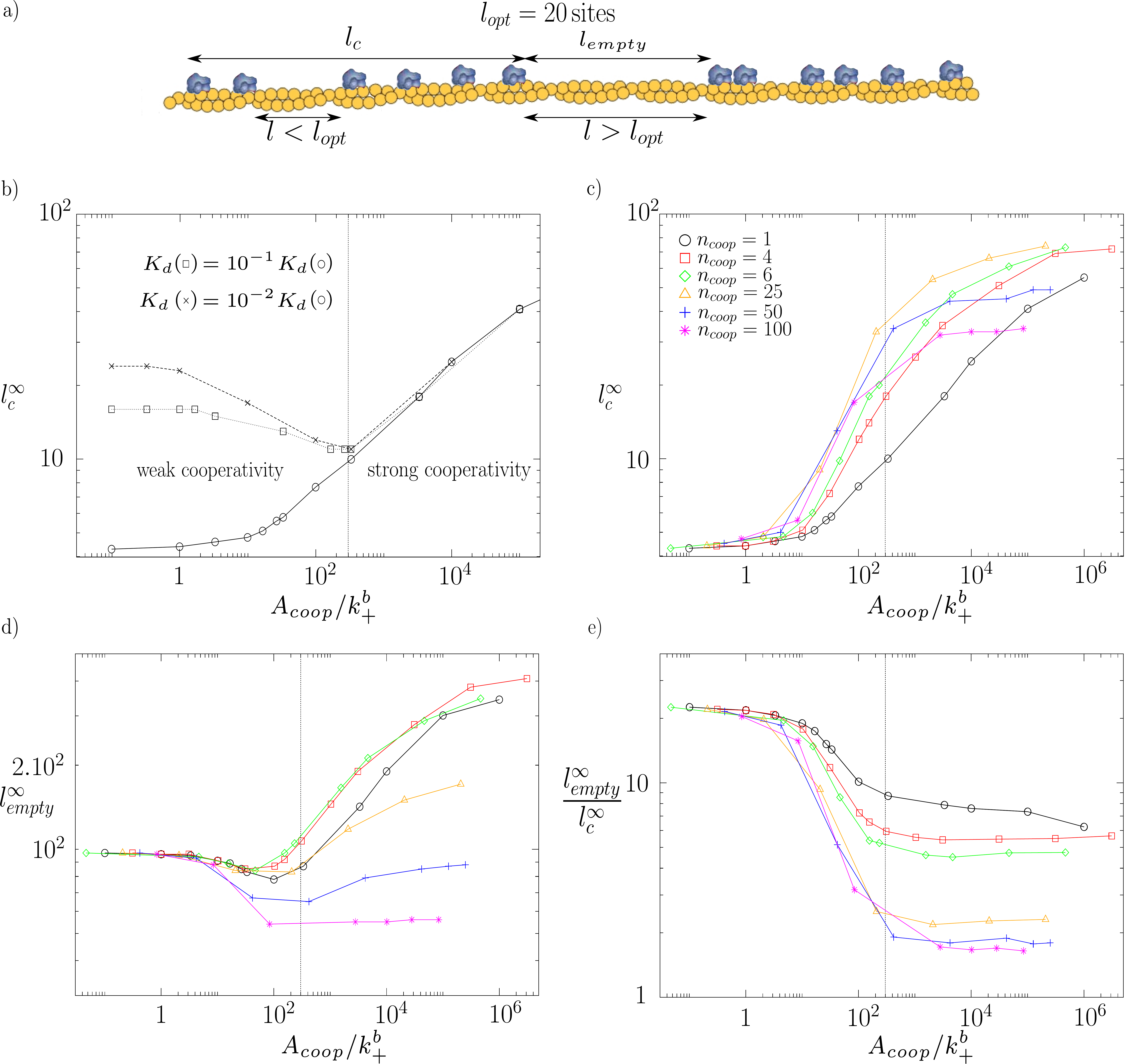}
\caption{Evolution of the steady state length of ABP clusters $l_c^{\infty}$ and distance between clusters $l_{empty}^{\infty}$ for different cooperative behaviors. Values of $l_c^{\infty}$, $l_{empty}^{\infty}$ and $n_{coop}$ are given in number of adhesion sites. a) Schematic cartoon defining the different length notations used in this study. The $\infty$ symbol indicates steady state values. b) Average cluster length $l_c^{\infty}$ as a fonction of the amplitude of cooperativity $A_{coop}/k_+^b$ for different values of affinity $K_d$ and for a range of cooperativity  $n_{coop}=1$. The $K_d$ represented by open circles is equal to $2.2.10^{-5}~M$ (value used in c), d) and e)). The vertical dashed line delimits the weak cooperativity regime from the strong cooperativity regime. c) Average cluster length $l_c^{\infty}$ as a fonction of the amplitude of cooperativity $A_{coop}/k_+^b$ for different ranges of cooperativity $n_{coop}$. d) Average distance between clusters $l_{empty}^{\infty}$ for different ranges of cooperativity $n_{coop}$ e) Ratio between the average distance between clusters $l_{empty}^{\infty}$ and the average cluster length $l_c^{\infty}$ as a fonction of the amplitude of cooperativity $A_{coop}/k_+^b$.}
\label{fig:3}
\end{figure}

The use of the model to understand how clusters of ABPs are formed required first a precise definition of clusters. Defining clusters as strictly consecutive bound ABPs along an actin filament was not very informative, as many clusters presented empty gaps due to the natural stochasticity of the process. This strict definition also did not permit us to compare the results of the simulations with experimental results obtained by fluorescence microscopy. In such assays, the limit of detection between 2 close molecules along an actin filament is limited by diffraction. The minimal distance to distinguish 2 fluorescent molecules emitting simultaneously is determined by the Rayleigh criterion, which corresponds to a distance of about 200 nm for usual optical wavelengths. Better resolutions can be achieved with super-resolution methods which typically decrease the limit of separation to distances of few tens of nanometers, corresponding to an approximate distance $l_{opt} = 20$ actin subunits (\cite{deschout2014} and our unpublished data). We used this reasonable criterion as a standard length to separate individual clusters in our model (Fig. \ref{fig:3}-a).

A large range of concentration of ABPs was tested (see Fig. \ref{fig:si1}). Similar to what is observed in vitro by fluorescence microscopy \cite{gressin2015}, low concentrations of ABPs triggered the formation of small and localized clusters of ABPs along the actin filament, while larger concentrations induced a quite abrupt transition to a fully decorated state. We then aimed at determining the influence of the amplitude of cooperativity $A_{coop}$ for the average cluster length $l_c^{\infty}$ at steady state labeled by the index $\infty$ (see Fig. \ref{fig:3}-b). The number of ABPs was fixed to 5000, which corresponds to a concentration of about $370~nM$. We started studying the effect of changing the affinity $K_d$ of the ABP for a fixed range of cooperativity $n_{coop}=1$ (Fig. \ref{fig:3}-b)). Two regimes can be identified. For an amplitude of cooperativity $A_{coop}/k_+^b$ lower than about 300, the affinity $K_d$ has a dramatic effect on the average cluster length $l_c^{\infty}$. We identify this regime as a weak cooperativity regime because in the limit of very small $A_{coop}$, the behavior is dominated by non-cooperative binding. In this extreme case, the ABPs bind to the filament randomly in space with approximately the same rates they would have on a bare filament. Moreover, the number of bound ABPs increases for smaller dissociation constants, increasing the probability to form long clusters. In the weak cooperativity regime, the length of the clusters $l_c$ is therefore mainly ruled by the affinity of the ABP $K_d$.  For $A_{coop}/k_+^b$ higher than about 300, the behavior becomes dominated by cooperative binding. We identify this regime as a strong cooperativity regime because the average cluster length $l_c^{\infty}$ at steady state does not depend on $K_d$. The total amount of bound ABPs and their distribution along the filament is also independent of $K_d$ (data not shown). The cluster length $l_c$ is not led by probability, but rather by the amplitude of cooperativity $A_{coop}$. The results are similar for longer range cooperativity: two regimes are observed with a boundary at the nearly same values (data not shown).

We can relate from Eq.\eqref{eqkcoop} the probability to bind the adjacent sites of a bound ABP with the characteristics of cooperativity for a given ABP corresponding to the boundary. $A_{coop}/k_+^b=300$ means that if $n_{coop}=1$, an ABP will bind next to a single bound ABP 301 times faster than to a bare filament. If $n_{coop}=100$, it will bind next to a single bound ABP about 4.6 times faster than to a bare filament. In both situations, an ABP will bind next to a long decorated portion 316 times faster compared to a bare filament. For proteins like tropomyosin, where $n_{coop}$ is believed to be equal to 1, values reported in literature indicate an increase of the binding probability of $\simeq10^2-10^3$ times to the sites that are adjacent to an already bound tropomyosin compared to a random site along the filament \cite{wegner1979,hill1992,vilfan2001}. These values are sufficient to characterize the parameters of cooperativity, and suggest that tropomyosin is at the boundary between the weak and strong cooperativity regimes. However, for ADF/cofilin, reported values indicate a 2.3 times faster binding rate in the vicinity of an already bound cofilin \cite{hayakawa2014}, but the uncertainty in the range of cooperativity $n_{coop}$ does not permit to evaluate unambiguously the amplitude of cooperativity $A_{coop}$.

We then aimed at determining the influence of the range of cooperativity for the average cluster length at steady state $l_c^{\infty}$. We fixed in these simulations a value of $K_d = 2.2.10^{-5}~M$ and evaluated the effect of modulating the range of cooperativity $n_{coop}$ (Fig. \ref{fig:3}-c)). The behavior for small $A_{coop}$ is mainly ruled by the affinity $K_d$ and not by cooperativity, and explains the formation of very short clusters. We will therefore mainly focus on the strong cooperativity regime. The first observation is that the cluster length $l_c^{\infty}$ increases with the amplitude of cooperativity $A_{coop}$. For $10^2<A_{coop}/k_+^b<10^5$, $l_c^{\infty}$ increases with the range of cooperativity $n_{coop}$ until $n_{coop}$ = 25 binding sites, but decreases for values of $n_{coop}$ greater than 25 binding sites. This means that there is an optimum range of cooperativity to form long clusters. In the limit of short range cooperativity, ABPs will be progressively recruited to the nearest sites available and form short clusters. In the opposite limit of long range cooperativity, ABPs will have the opportunity to bind further away along the actin filament, which will increase the likelihood to form new independent clusters instead of growing the original cluster. A second observation is that $l_c^{\infty}$ seems to reach a plateau in the limit of very strong cooperativity, and the asymptotic values decrease with increasing values of $n_{coop}$. Some additional points would be necessary for large values of $A_{coop}$ to confirm rigorously this assertion, but very large values of $A_{coop}$ are probably unrealistic and would require very long simulation times.

In addition to cluster length, we noticed that cooperativity also modulates the distribution of clusters along actin filaments. To study this effect, we calculated $l_{empty}^{\infty}$, which is the average distance between two consecutive clusters at steady state, as a function of $A_{coop}/k_+^b$ (Fig. \ref{fig:3}-d)). As for cluster length $l_c^{\infty}$, $l_{empty}^{\infty}$ does not change with $n_{coop}$ for small $A_{coop}$ where ABP binding is mainly ruled by the affinity $K_d$ and not by cooperativity. For $10^2<A_{coop}/k_+^b<10^5$, there is an optimal range of cooperativity $n_{coop} = 4-6$ for which clusters are most distant from one another. Moreover, $l_{empty}^{\infty}$ increases in the strong cooperativity regime until it probably reaches a plateau under the limit of very strong cooperativity. This result indicates that the formation of long clusters is correlated with the maintenance of long sections of bare filaments between clusters. In this regime, it is more probable to form long clusters than to create new ones. 
The ratio between the distance between consecutive clusters and their length remains rather constant in the strong cooperativity regime (Fig.\ref{fig:3}-e)). This asymptotic value decreases with the range of cooperativity $n_{coop}$ from about 6-7 for $n_{coop}=1$ to about 2 for $n_{coop}=100$. In the case of a long range cooperativity, $l_c^{\infty}$ and $l_{empty}^{\infty}$ are both quite small and of similar values (about 35 and 55 adhesion sites respectively). This result indicates that a long range cooperativity is unable to form isolated clusters with the current definition, and cannot account for observations made by fluorescence microscopy \cite{gressin2015}. 

To summarize this section, our model predicts that for cooperativity amplitudes $10^2<A_{coop}/k_+^b<10^5$, different optimum ranges of cooperativity $n_{coop}$ are predicted to obtain specific characteristics of clusters. A range of cooperativity of $n_{coop}=25$ is optimal to form long clusters of ABPs, while a range of cooperativity of $n_{coop}=4-6$ is optimal to form distant clusters of ABPs. However, it is for the shortest ranges of cooperativity ($n_{coop}=1$) that the distance between cluster is the longest with respect to their length. On the contrary, very long range cooperativities are ineffective to form isolated clusters.

\subsection*{Impact of Cooperativity on the Number of ABPs per Cluster}

As we previously defined a length $l_{opt} = 20$ empty adhesion sites above which clusters are optically separated, the length of the clusters is not correlated with the number of ABPs per cluster. In this section, we focus on the impact of cooperativity in modulating the ABP density within individual clusters at steady state.

\begin{figure}[hbt!]
\centering
\includegraphics[width=1\linewidth]{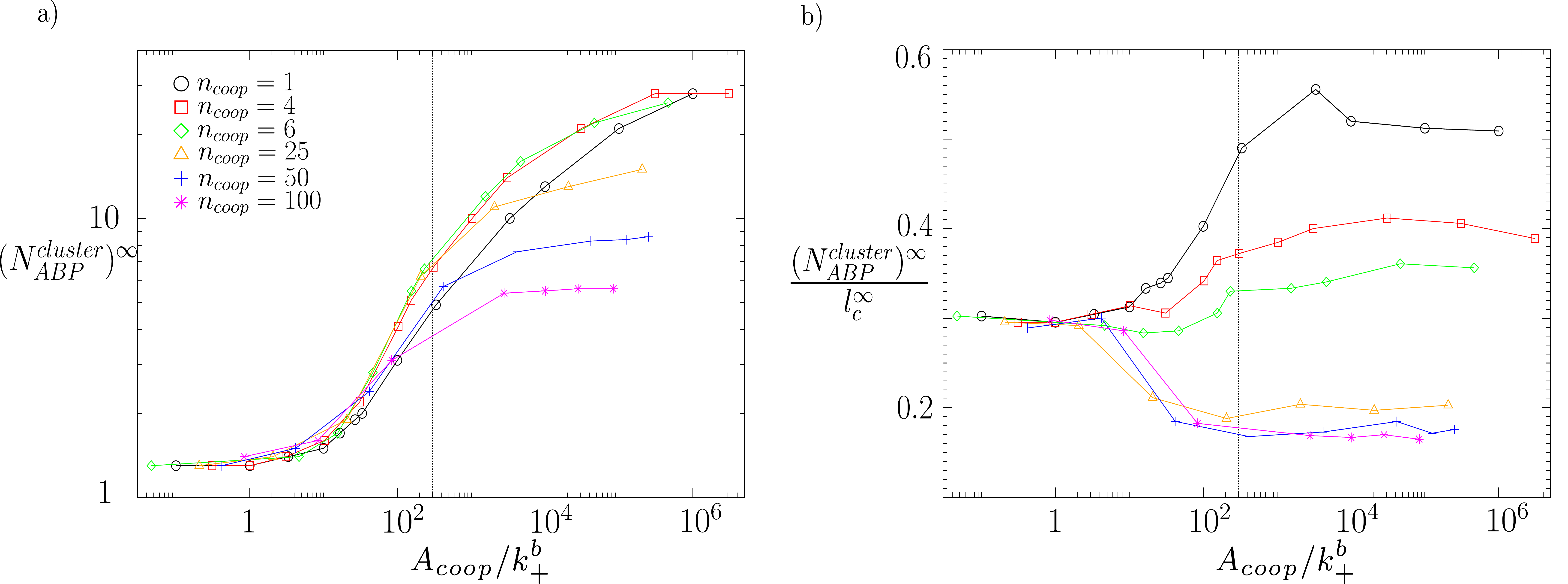}
\caption{a) Evolution of the steady state number of ABPs per cluster $(N_{ABP}^{cluster})^{\infty}$ as a fonction of the amplitude of cooperativity $A_{coop}/k_+^b$ for different ranges of cooperativity $n_{coop}$. b) Density of ABPs per cluster $(N_{ABP}^{cluster})^{\infty}/l_c^{\infty}$ as a fonction of the amplitude of cooperativity $A_{coop}/k_+^b$ for different ranges of cooperativity $n_{coop}$. The dashed lines indicate the boundary between weak and strong cooperativity regimes.}
\label{fig:4}
\end{figure}

We first investigated how the amplitude of cooperativity $A_{coop}/k_+^b$ influences the number of ABPs per cluster $N_{ABP}^{cluster}$ (Fig. \ref{fig:4}-a)). In the strong cooperativity regime, $(N_{ABP}^{cluster})^{\infty}$ increases with $A_{coop}/k_+^b$ until it reaches an asymptotic value which decreases with $n_{coop}$ (as mentioned previously, additional simulations would be necessary to confirm formally the existence of the plateaus). The typical maximum values are below 30 bound ABPs, which is much smaller than total number of 5000 ABPs; therefore, plateaus are not due to finite-size effects. For $10^2<A_{coop}/k_+^b<10^3$, the number of ABPs per cluster is maximum for a range of cooperativity $n_{coop}$ between 6 and 25. The non monotonous evolution of $(N_{ABP}^{cluster})^{\infty}$ as a function of $n_{coop}$ indicates an optimal range of cooperativity to accumulate proteins within clusters. For larger $A_{coop}/k_+^b$, $(N_{ABP}^{cluster})^{\infty}$ saturates faster for long ranges of cooperativity. As a consequence, for $10^3<A_{coop}/k_+^b<10^6$, the optimum range of cooperativity $n_{coop}$ is lower, between 4 and 6. In contrast, the curve of $n_{coop}=1$ crosses over all the other curves. This indicates that end-to-end cooperativity is an efficient way to recruit a large number of ABPs per clusters, as long as the amplitude of cooperativity is strong enough (i.e. $10^6<A_{coop}/k_+^b$) to drive the process. Conversely, a long range cooperativity is ineffective to recruit many ABPs within one cluster at any $A_{coop}$. For example, the maximum $(N_{ABP}^{cluster})^{\infty}$ is 5.6 for $n_{coop}=100$.

We can derive the density of bound ABPs per cluster at steady state defined as the ratio between the number of ABPs per cluster $(N_{ABP}^{cluster})^{\infty}$ and the cluster length $l_c^{\infty}$ (Fig. \ref{fig:4}-b)). Strikingly, these curves evolve in opposite directions when approaching the strong cooperativity regime. While ABP density per cluster decreases  with $A_{coop}/k_+^b$ for long range cooperativity (i.e. $n_{coop}=25,50$ or 100), ABP density increases for short range cooperativity. The reason is that for large ranges of cooperativity $n_{coop}$, ABPs can be stochastically recruited at larger distances than the average distance between two consecutive bound ABPs within one cluster. As a consequence, the density declines. By contrast, for small $n_{coop}$, ABPs bind at distances which is shorter than the typical distance between ABPs within clusters, therefore the ABP density grows. Moreover, a plateau is reached at larger $A_{coop}/k_+^b$ for all ranges of cooperativity. The values of this plateau in the limit of very strong cooperativities decrease with $n_{coop}$, indicating that a high density of protein per cluster requires necessarily a short range of cooperativity. 

To summarize this section, our model predicts that for cooperativity amplitudes $A_{coop}/k_+^b<10^6$, a range of cooperativity of $n_{coop}=4-6$ is optimal to recruit many ABPs per cluster. Only for stronger cooperativities is the limit of very short range of cooperativity $n_{coop}=1$ better. Regarding the ABP density of these clusters, the limit of very short range cooperativity is systematically optimal. On the contrary, long range cooperativities are inefficient to recruit many ABPs per clusters and to form dense clusters.

\subsection*{Impact of Cooperativity on Clusters Rate of Assembly}

\begin{figure}[hbt!]
\centering
\includegraphics[width=1\linewidth]{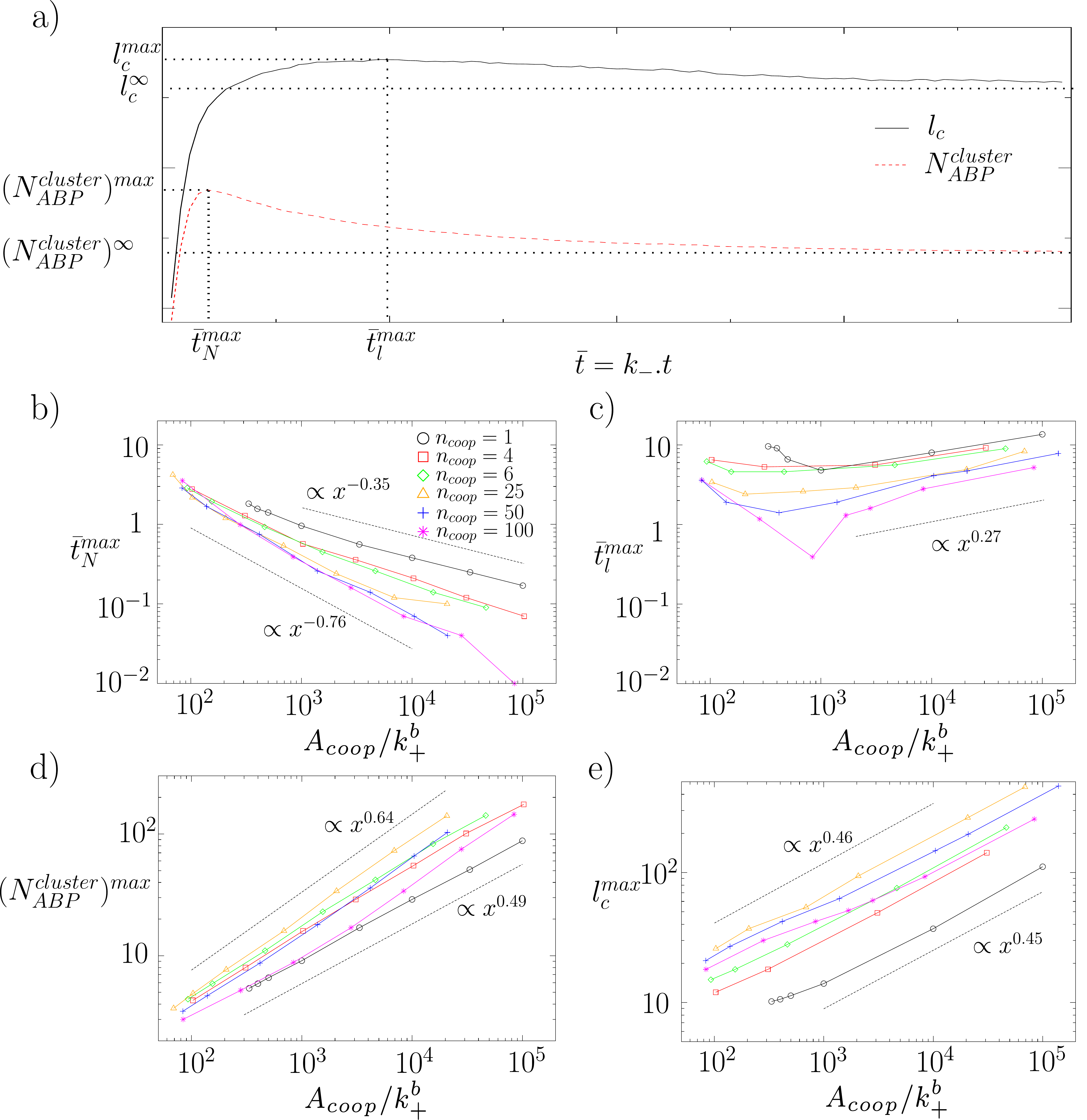}
\caption{Dynamics of cluster assembly. Time is expressed in number of $1/k_-$ and cluster length $l_c$ in number of binding sites. a) Evolution of average cluster length $l_c$ and average number of ABP per cluster $N_{ABP}^{cluster}$ over time. b) Average time $\bar{t}_N^{max}$ for clusters to reach the highest number of bound proteins $(N_{ABP}^{cluster})^{max}$ as a function of the amplitude of cooperativity $A_{coop}/k_+^b$. c) Average time $\bar{t}_l^{max}$ for clusters to reach the longest size $l_c^{max}$ as a function of the amplitude of cooperativity $A_{coop}/k_+^b$. d) Maximum number of bound proteins $(N_{ABP}^{cluster})^{max}$ as a function of the amplitude of cooperativity. e) Maximum length of the clusters $(N_{ABP}^{cluster})^{max}$ as a function of the amplitude of cooperativity. In b,c,d,e), the dotted black lines represent power functions with exponents corresponding to the fits of the surrounding curves.}
\label{fig:5}
\end{figure}

While results presented in the previous sections correspond to cluster configurations at steady state, this section focuses on their dynamics of assembly. A dimensionless time $\bar{t}$ is expressed in number of $1/k_-$ and all ABPs are unbound at $\bar{t}=0$. In the case of ADF/cofilin, $k_-$ is reported to be $5.10^{-3}-5.10^{-1}~s^{-1}$ \cite{blanchoin1999, wioland2017}, so the unit for $\bar{t}$ is $2-200~s$. For tropomyosin, the dynamics is faster as $k_-=10^2-10^3~s^{-1}$ \cite{weigt1991,vilfan2001,christensen2017}, leading to time unit of $10^{-3}-10^{-2}~s$. Nevertheless, the end-to-end cooperativity may decrease the unbinding rate of one or two orders of magnitude for tropomyosin. Additionally, to avoid the noise of early dynamics due to finite size effects, we simulated a larger system of 40000 ABPs for 640000 adhesion sites, while keeping the same concentrations and dissociation constants as in previous sections. 

We analyzed the growth and dynamics of ABPs clusters from the binding of the first molecules until steady state. The total amount of bound ABPs increases smoothly until equilibrium amount is attained (data not shown). In the meantime, both average length of clusters $l_c$ and average number of ABPs per cluster $N_{ABP}^{cluster}$ also increase with similar timescales. Interestingly, a maximum value appears for sufficiently large $A_{coop}/k_+^b$. This means they then decrease slightly toward their steady state values (see Fig. \ref{fig:5}-a)). The reason is that at early time and for large amplitude of cooperativity, the binding in vicinity of existing clusters is much faster than unbinding. As a consequence, almost all binding events contribute to amplify $N_{ABP}^{cluster}$ (and $l_c$) as long as available reservoir of ABPs is large enough. When this reservoir becomes limited, meaning that equilibrium amount of bound ABPs is attained, the number of unbinding events is no longer negligible and the dynamics of clusters is ruled by the substitution of bound ABPs rather than the addition of new ones. The likelihood that individual clusters split into two different clusters of smaller size is more critical and implies decrease of both $N_{ABP}^{cluster}$ and $l_c$.

Interestingly, the time $\bar{t}_N^{max}$ to reach the maximum number of proteins per cluster $(N_{ABP}^{cluster})^{max}$, when existing, is systematically shorter than the time $\bar{t}_l^{max}$ to reach the maximum length of clusters $l_c^{max}$ (see Fig. \ref{fig:5}-a,b,c)). The reason is that at $\bar{t}_N^{max}$, clusters are dense, particularly for small values of $n_{coop}$ (see Fig. \ref{fig:si2}), and only few adhesion sites are available except on the side of the clusters. Meanwhile, unbinding events continue to occur within clusters, diminishing $(N_{ABP}^{cluster})$. Moreover, as a consequence of cooperativity, the binding rate in the vicinity of clusters is large. Therefore, additional ABPs will bind preferentially outside clusters and extend them, which means that the average length of clusters $l_c$ continues to increase. As a consequence, $\bar{t}_l^{max}\geq\bar{t}_N^{max}$. It is also of interest to mention that typical distances separating clusters $l_{empty}$ are larger than every $n_{coop}$ used in this study for large $A_{coop}/k_+^b$ (see Fig. \ref{fig:si2}). This means that probability to merge two successive clusters is quite small.

We analyzed then how the parameters of cooperativity impact rates of cluster assembly. We first observed that the average time needed for clusters to reach their maximal number of ABP $\bar{t}_N^{max}$ decreases with the amplitude of cooperativity $A_{coop}/k_+^b$ following power laws (see Fig. \ref{fig:5}-b)). The reduction is explained by the fact that although clusters reach a higher number of bound ABP when $A_{coop}/k_+^b$ increases, the binding rate of ABPs increases faster with the amplitude of cooperativity. Furthermore, the time $\bar{t}_N^{max}$ diminishes with the range of cooperativity $n_{coop}$. The reason is that for short ranges of cooperativity, recruitment rates of new ABPs are slower because limited to lower numbers of binding sites. Nevertheless, the maximum number of protein per cluster $(N_{ABP}^{cluster})^{max}$ increases for $n_{coop}\leq25$ (see Fig. \ref{fig:5}-d)). The effect of $n_{coop}$ is therefore non intuitive as smaller clusters could be expected to assemble faster. This is not the case, and we can conclude from these results that increasing the range of cooperativity up to 25 subunits along an actin filament is a good strategy to have simultaneously a fast clustering and a large number of bound ABPs. In contrast, both the maximum number of bound ABP $(N_{ABP}^{cluster})^{max}$ and the average time needed to reach this threashold decline for $n_{coop}\geq25$. Overall, the optimal value to form populated clusters at steady state is $n_{coop}=4-6$ while it is $n_{coop}=25$ at $\bar{t}_N^{max}$ (see Fig. \ref{fig:4}-a) and Fig. \ref{fig:5}-d)), although clusters are less dense for $n_{coop}=25$ than for $n_{coop}=4-6$ at short time, when unbinding is negligible (see Fig. \ref{fig:si2}).

For the smallest amplitudes of cooperativity $A_{coop}/k_+^b$ in regime where maximums exist, the average time required to reach the maximal length of clusters $\bar{t}_l^{max}$ follows the same trend than the average time required to reach the maximal number of ABP per cluster $\bar{t}_N^{max}$. However, when $A_{coop}/k_+^b$ increases, $\bar{t}_l^{max}$ progressively diverges to eventually increase and follow a power law (see Fig. \ref{fig:5}-b,c)). As previously explained, when the equilibrium amount of bound ABPs is attained, the elongation of clusters at their extremities is mostly due to the substitution than to the addition of new ABPs. Therefore, when $A_{coop}/k_+^b$ increases, the binding rate of ABP is more important in the vicinity of the clusters than in bare sections of the filament, and new binding events enlarge existing clusters. This is consistent with an increase of cluster density at $\bar{t}_l^{max}$ with $A_{coop}/k_+^b$ (see Fig. \ref{fig:si2}). Restriction of new binding area slows down elongation and increases $\bar{t}_l^{max}$. This effect is even stronger for short range cooperativities where this restriction is more important. As a result, $\bar{t}_l^{max}$ increases with $A_{coop}/k_+^b$ for strong cooperativity and decreases with the range of cooperativity $n_{coop}$. Finally, the maximum cluster length $l_c^{max}$ grows with the amplitude of cooperativity $A_{coop}/k_+^b$ following a power law with an exponent which is independent of the range of cooperativity $n_{coop}$ (Fig. \ref{fig:5}-e). This exponent is smaller than for $N_{ABP}^{cluster}$ (Fig. \ref{fig:5}-d), leading to denser clusters in the strong cooperativity limit. Moreover, similar to the steady state value $l_c^{\infty}$, the maximum cluster length $l_c^{max}$ is optimal for $n_{coop}=25$. Though, inverse to $l_c^{\infty}$, the maximum value $l_c^{max}$ does not seem to saturate in limit of very strong cooperativity. This absence of saturation is also observed for $N_{ABP}^{cluster}$.

To summarize this part, the assembly rate of the ABP clusters is correlated with the range of cooperativity $n_{coop}$. The shorter is the range, the slower is the dynamics. Then, a maximum value appears for $l_c$ and $N_{ABP}^{cluster}$ when $A_{coop}/k_+^b$ is large enough. The time required to reach the maximum number of ABPs per clusters decreases with the amplitude of cooperativity $A_{coop}/k_+^b$, while the time to reach the maximum cluster length increases for strong cooperativity.

\subsection*{System Configurations at Fixed Number of ABPs per Clusters}

\begin{figure}[hbt!]
\centering
\includegraphics[width=1\linewidth]{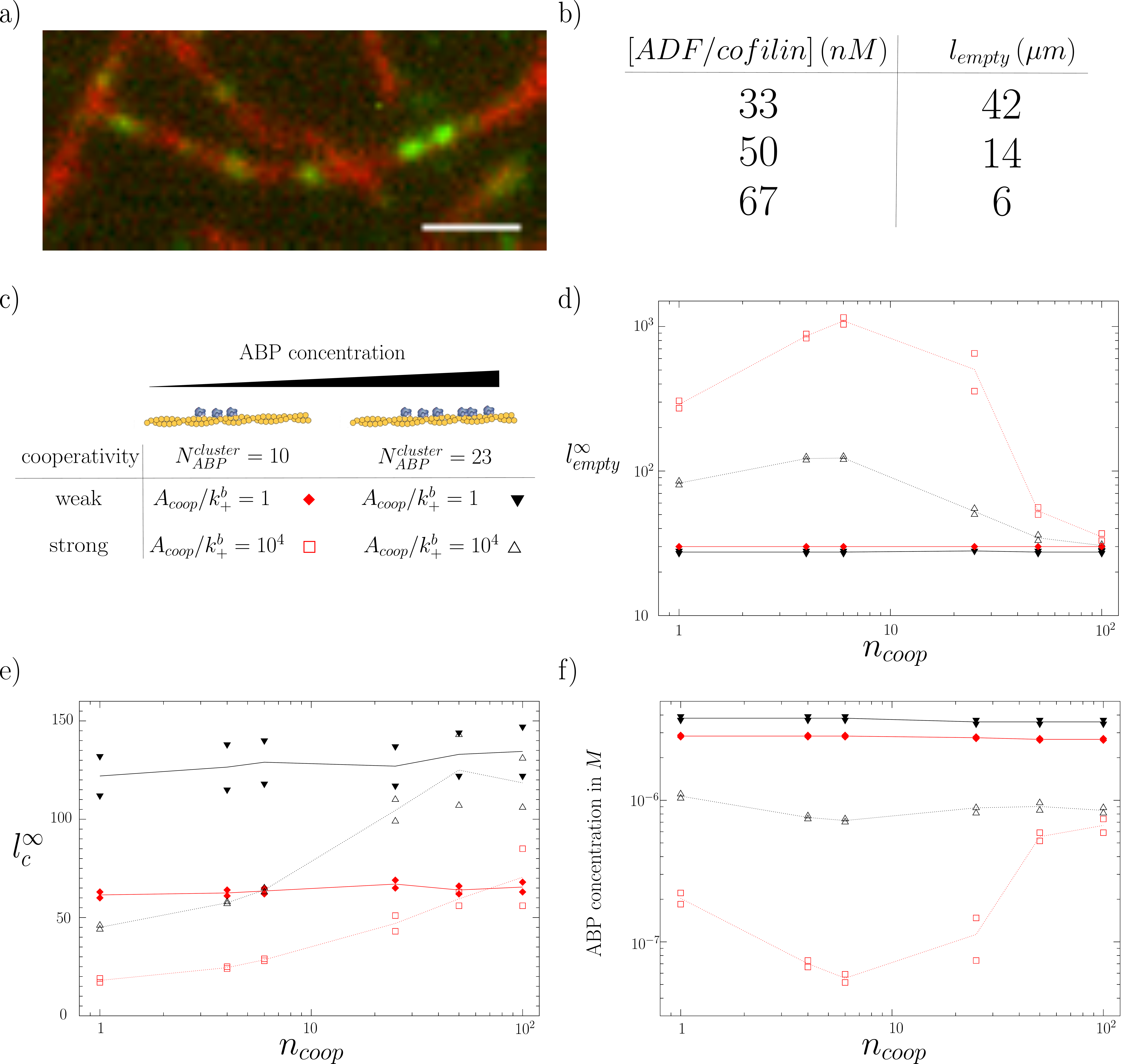}
\caption{Predicted configurations of the system to generate clusters of 10 and 23 ABPs on average at steady state. Values of $l_{empty}^{\infty}$, $l_c^{\infty}$ and $n_{coop}$ are given in number of adhesion sites. Each datapoint indicates the values obtained from the closest simulations. Curves link average values. a) Example of multiple Alexa-488-labeled ADF/cofilin (67 nM; in green) clusters formed on the side of an individual actin filament (in red). Scale bar=3$~\mu m$. b) Experimental values measured from a). c) Cartoon explaining the significance of symbols. Closed symbols (resp. opened) are results of simulations for the weak (resp. strong) cooperativity regime. d) Average distance between successive clusters $l_{empty}^{\infty}$ as a function of the range of cooperativity $n_{coop}$. e) Average cluster length $l_c^{\infty}$ as a function of the range of cooperativity $n_{coop}$. f) Average concentration of ABP required to obtain clusters of 10 and 23 molecules as a function of the range of cooperativity $n_{coop}$.}
\label{fig:6}
\end{figure}

We now focus on a well-described family of proteins, ADF/cofilins, as an example of how the model can be used to predict the characteristics of cooperativity of an ABP. The cooperativity of ADF/cofilin binding to actin filaments has been reported extensively, but the measured amplitude and range of cooperativity along actin filaments varies considerably between studies. Microscopy experiments previously revealed the behavior of ADF/cofilin binding to individual actin filaments. Above a minimal concentration of ADF/cofilin in solution, the protein binds visibly to actin filaments and forms stable clusters of reproductible size at the 10 seconds timescale \cite{gressin2015}. We measured experimentally for this study the average distance between clusters at three different concentrations of ADF/cofilin (Fig. \ref{fig:6}-a,b) and Fig. \ref{fig:si3}). Measured values vary from 6 to 42 $\mu$m, which corresponds to 2000 to 14000 subunits of actin. The order of magnitude between experimental measurements and the model are comparable. Moreover, the trend is similar, as the simulations predict that $l_{empty}$ decreases with the concentration of ABPs (see Fig. \ref{fig:si1}-c,d)).

Then the number of bound molecules was evaluated by quantitative fluorescence microscopy, and was found to remain below 10 molecules per cluster for a concentration of 180 nM of ADF/cofilin, and to reach on average 23 molecules per cluster for a concentration of 360 nM of ADF/cofilin. When the threshold of 23 molecules is reached, these isolated clusters are triggering the polarized severing of the actin filaments, towards their pointed ends, at the interface between bare and decorated sections of the filament. At higher concentration of ADF/cofilin, the protein fully decorates, copolymerizes with actin and stabilizes the filaments \cite{gressin2015}. While previous simulations were run with a constant pool of ABPs, we varied in the new simulations the pool of ABPs. We determined the cooperativity conditions leading to the formations of experimentally observed clusters of 10 or 23 bound proteins on average, and analyzed for these configurations the characteristics of the clusters (Fig. \ref{fig:6}). We generated results for 2 representative amplitudes of cooperativity $A_{coop}/k_+^b=1$ and $10^4$, which corresponded respectively to the weak and strong cooperativity regimes. 

A first general observation is that all values are weakly influenced by the range of cooperativity $n_{coop}$ in the weak cooperativity regime. The length of clusters in this regime is always longer than the distance $l_{empty}$ separating them. This means that clustering observed in experiments with isolated clusters of ADF/cofilin is not compatible with the weak cooperativity regime. In contrast, in strong cooperativity regime, the range of cooperativity $n_{coop}$ has a strong influence in case of 10 ABPs per cluster. The average distance between clusters increases from about 300 adhesion sites between clusters for $n_{coop}=1$ to about 1000 adhesion sites (the order of magnitude measured experimentally) for the optimum value $n_{coop}=6$. Above $n_{coop}=6$, it decreases progressively to reach the same level than the weak cooperativity regime for the longest ranges of cooperativity (Fig. \ref{fig:6}-d). Evolution for clusters of 23 ABPs is similar with smaller $l_{empty}$ (about 100 adhesion sites for maximum value). These results exclude in the case of ADF/cofilin the possibility of a range of cooperativity much above 25 actin subunits, and makes a very short range of cooperativity much less likely than values around $n_{coop}=4-6$.

Another important experimental observation is that the size of ADF/cofilin clusters remains rather small \cite{gressin2015}. They systematically appear as spots by fluorescence microscopy, which strongly suggests that all molecules are not distinguishable optically and located within a section of less than about 200 nm, which corresponds to about 20 adhesion sites. In the weak cooperativity regime, the average cluster length $l_c^{\infty}$ is systematically greater than 70 adhesion sites (Fig. \ref{fig:6}-e). In the strong cooperativity regime, our simulations indicate on the contrary that below a range of cooperativity $n_{coop}=10$, clusters of ABPs are short enough to account for the experimental data in case of 10 ABPs per cluster. Nevertheless, for cluster of 23 ABPs, simulations values seem a little bit overestimated. This result suggests again that the larger ranges of cooperativity are less probable in the case of ADF/cofilin.

We also investigated the concentration of ABPs which is necessary to form these clusters (Fig. \ref{fig:6}-f). As mentioned above, in weak cooperativity regime, the formation of clusters is independent of the concentration of ABPs. The typical concentrations required to form clusters in this regime are 2.8 and 3.8$~\mu$M respectively for small and large clusters (or about half the actin concentration), which is well above what is needed \textit{in vitro}. The required amount of protein is smaller in the strong cooperativity regime, and the effect is non monotonous with the range of cooperativity $n_{coop}$. The smallest required amount of ADF/cofilin is predicted for a range of cooperativity of $n_{coop}=6$. For this range of cooperativity, clusters of 23 ABPs are formed at a concentration of $[ADF/cofilin]\simeq0.7~\mu M$ ($\simeq0.1[actin]$), and clusters of 10 ABPs are formed at a concentration of $[ADF/cofilin]\simeq50~nM$ ($\simeq0.008[actin]$). These concentrations are in relatively good agreement with concentrations used in the experiments. Typical time to reach maximum number of ABPs per cluster at $A_{coop}/k_+^b=10^4$ and $n_{coop}=6$ is $0.2/k_-$. For cofilin, $k_-=5.10^{-3}-5.10^{-1}~s^{-1}$ \cite{blanchoin1999, wioland2017}, so typical timescales are $0.4-40~s$. This is in agreement with the time needed to reach clusters of 23 cofilin in experiments \cite{gressin2015}.

To summarize this part, evolution of $l_{empty}$ in simulations is consistent with experiments even if strict comparison is difficult here. Then, the model suggest that in case of ADF/cofilin, long range and/or weak cooperativity unlikely.

\section*{Conclusion}
\begin{figure}[hbt!]
\centering
\includegraphics[width=1\linewidth]{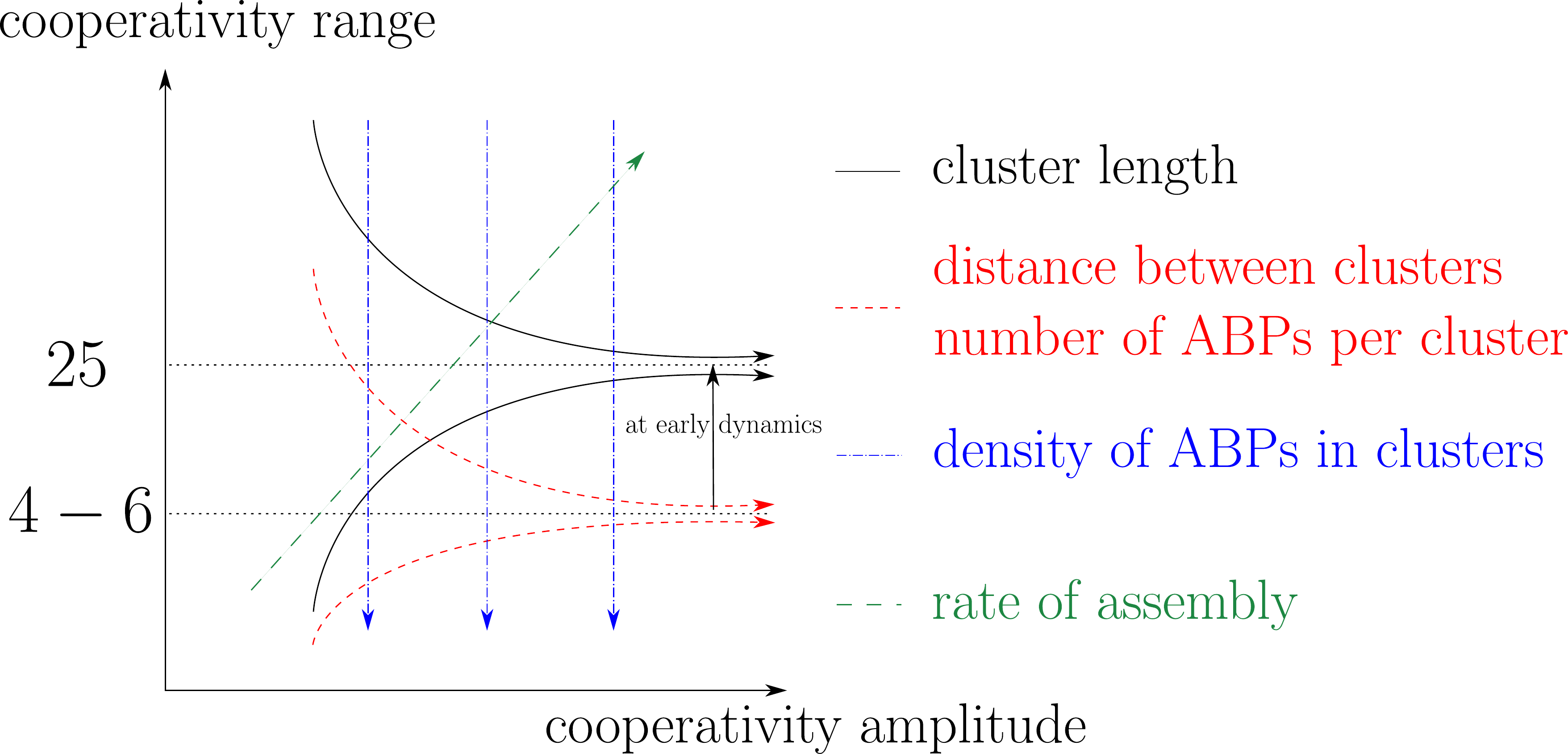}
\caption{Cartoon summarizing how the amplitude and range of cooperativity impact ABP cluster formation. Each arrow indicates how specific properties of cluster formation are enhanced from their lower values to their higher values in this 2-dimensional diagram.}
\label{fig:7}
\end{figure}

In this study, we have detailed the impact of binding cooperativity for the formation of protein clusters to the side of individual actin filaments. We have principally investigated the influence of two parameters which are the amplitude and the range of cooperativity. We found that both parameters have a strong influence on cluster assembly properties such as size, density and dynamics of the clusters, and these effects are summarized in Fig. \ref{fig:7}. These properties are impacted differently by the parameters of cooperativity. For example, building dense clusters of ABPs at a fast rate will require trade-off values of the amplitude and range of cooperativity. As ABPs have been demonstrated experimentally to follow a diversity of cooperative binding behaviors, it is expected that cooperativity is used in cells as a powerful mechanism to control the decoration of actin filaments. Depending on the function of the ABP and the level of activity required for a specific population of actin filaments, cells will favor locally the formation of clusters of appropriate size, density and dynamics.

\section*{Author Contributions}
T.L.G. and A.M. conceived the project and wrote the manuscript. T.L.G. designed the model, performed the simulations, analyzed the computational results and compared with the experimental results. A.M. performed and analyzed the experiments and funded the project.

\section*{Acknowledgments}
The authors would like to thank Christopher P. Toret for his critical reading of the manuscript. This project has received funding from the European Research Council (ERC) under the European Union's Horizon 2020 research and innovation programme (grant agreement $n\textsuperscript{o}$ 638376/Segregactin).

\newpage

\bibliographystyle{plain}
\bibliography{mainarxiv}

\newpage
\section*{Supplementary Material}

\setcounter{figure}{0} 
\renewcommand\thefigure{S\arabic{figure}} 

\begin{figure}[hbt!]
\centering
\includegraphics[width=0.9\linewidth]{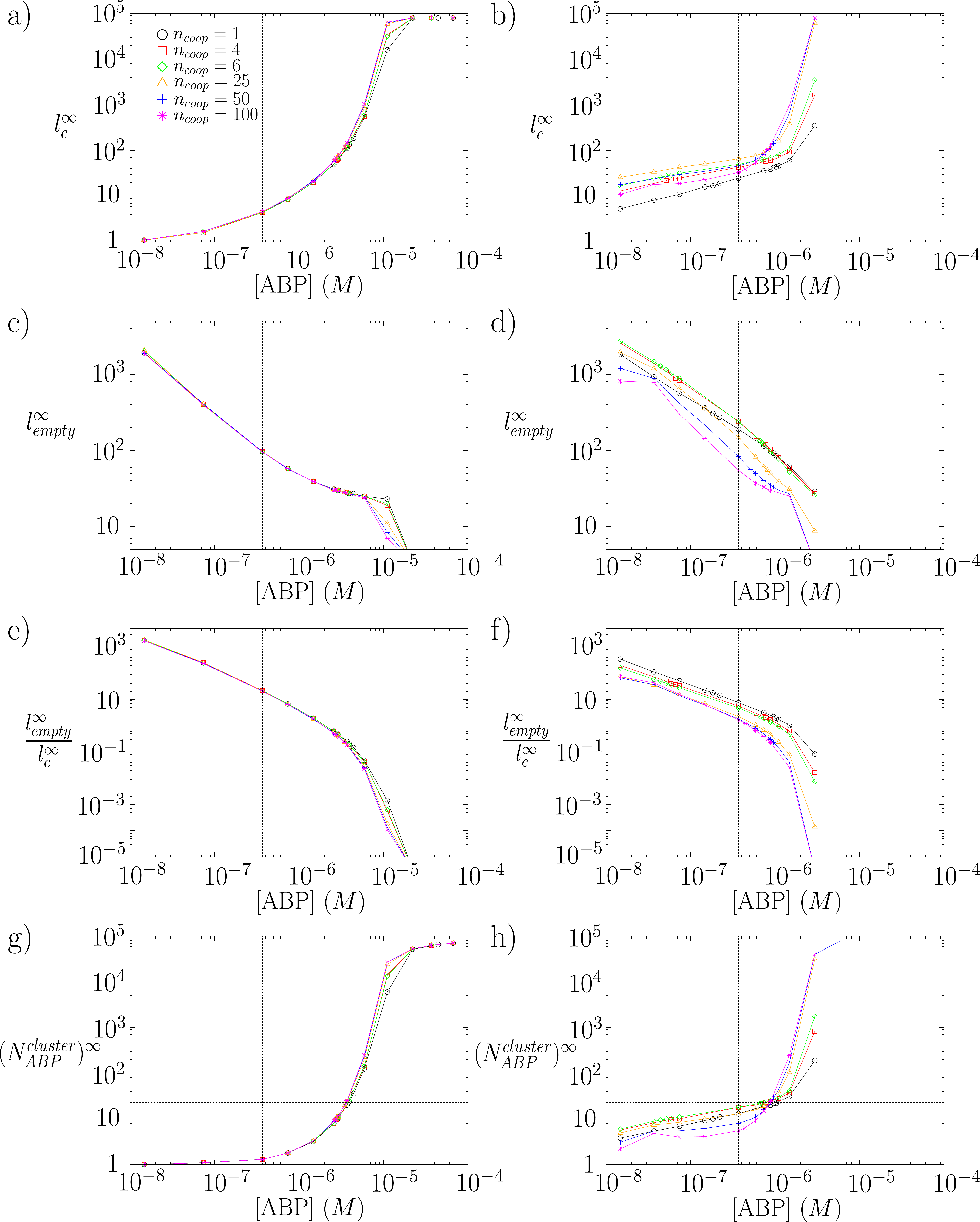}
%
\caption{Evolution of the steady state clusters characteristics (a) and b) average cluster length; c) and d) average distance between clusters ; e) and f) ratio between the distance between clusters $l_{empty}^{\infty}$ and the cluster length $l_c^{\infty}$; g) and h) number of ABP per cluster) as a function of the concentration of ABP. The filament length is $8.10^4$ binding sites. Left plots correspond to the weak cooperativity regime $A_{coop}/k_+^b=1$. Right plots correspond to the strong cooperativity regime $A_{coop}/k_+^b=10^4$. Vertical dotted lines represent conditions equivalent to 5000 and 80000 ABPs respectively in a volume of $2.25.10^7l_{site}^3$. The horizontal dotted lines in g) and h) correspond to clusters of 10 and 23 molecules.}
\label{fig:si1}
\end{figure}

\begin{figure}[hbt!]
\centering
\includegraphics[width=0.9\linewidth]{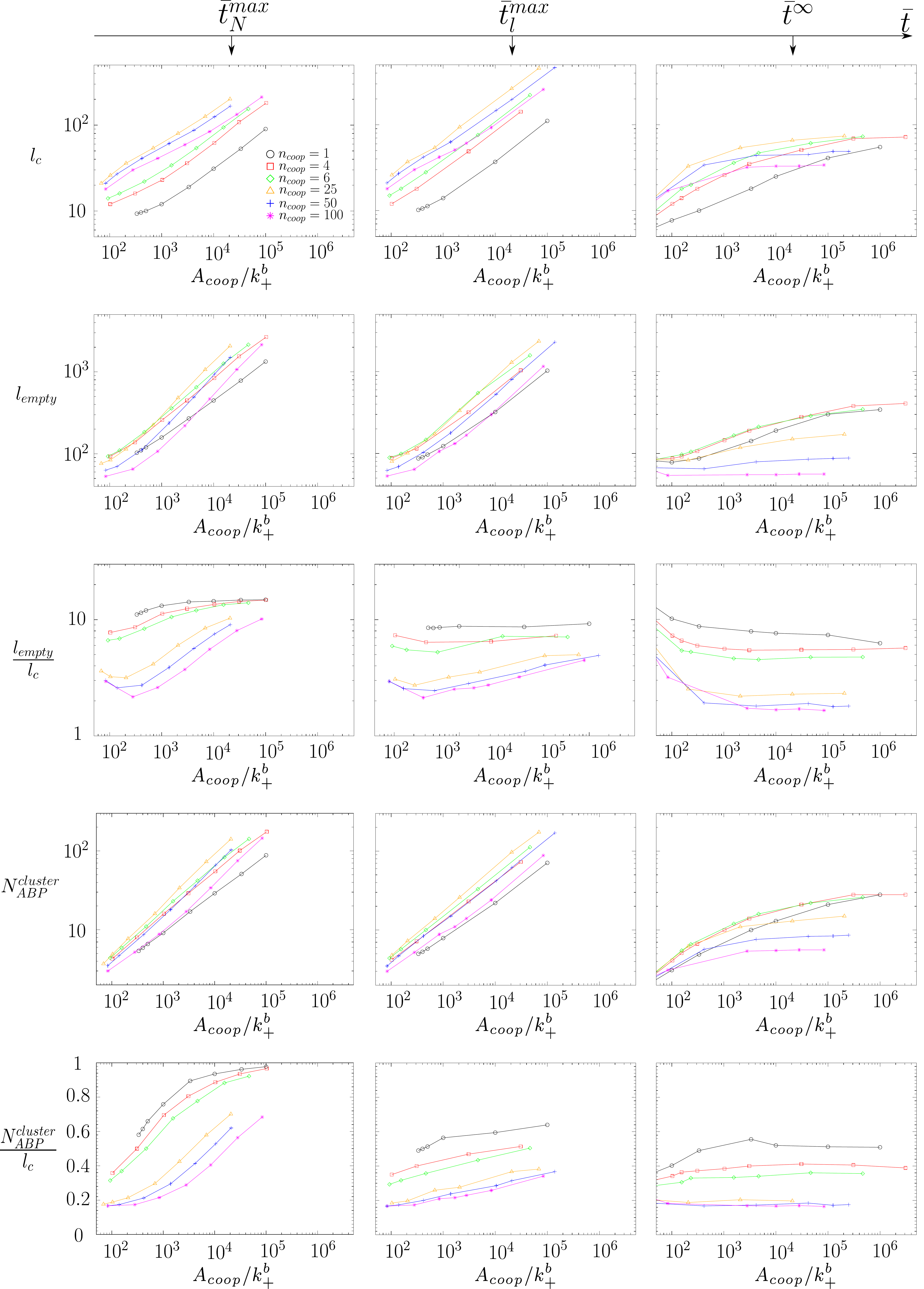}
\caption{Evolution of clusters characteristics (cluster length, distance between clusters, ratio between the distance between clusters and the cluster length, number of ABP per cluster and density of ABP per cluster) as a function of the amplitude of cooperativity $A_{coop}/k_+^b$. Left plots are data when system reach the highest number of proteins per clusters $(N_{ABP}^{cluster})^{max}$; center plots are data when clusters reach the longest size $l_c^{max}$; right plots are data at steady state.}
\label{fig:si2}
\end{figure}

\begin{figure}[hbt!]
\centering
\includegraphics[width=0.9\linewidth]{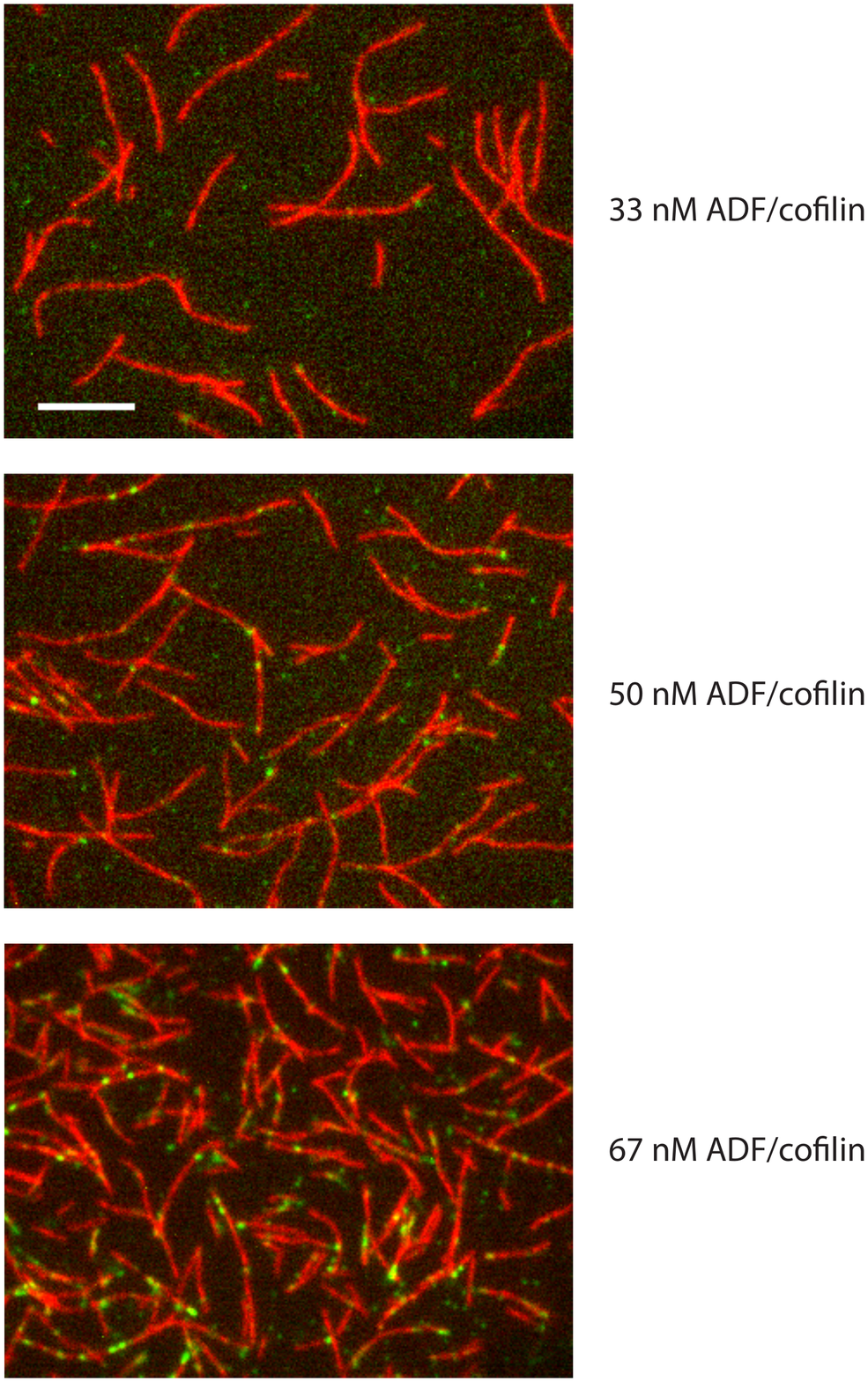}
\caption{Distribution of Alexa-488-labeled ADF/cofilin clusters (in green) along actin filaments (in red) for three concentrations of ADF/cofilin 20 min after the initiation of the experiment. Scale bar : 10$~\mu m$. Numbers of clusters per unit length of actin varies from 0.025 clusters per $\mu$m at 33 nM of ADF/cofilin, to 0.076 clusters per $\mu$m at 50 nM of ADF/cofilin and to 0.149 clusters per $\mu$m at 67 nM of ADF/cofilin.}
\label{fig:si3}
\end{figure}

\end{document}